\documentclass{SciPost}

\hypersetup{
    colorlinks,
    linkcolor={red!50!black},
    citecolor={blue!50!black},
    urlcolor={blue!80!black}
}

\usepackage[bitstream-charter]{mathdesign}
\usepackage{enumitem}
\usepackage{ulem}
\usepackage{amsmath}

\DeclareSymbolFont{usualmathcal}{OMS}{cmsy}{m}{n}
\DeclareSymbolFontAlphabet{\mathcal}{usualmathcal}

\fancypagestyle{SPstyle}{
\fancyhf{}
\lhead{\colorbox{scipostblue}{\bf \color{white} ~SciPost Physics }}
\rhead{{\bf \color{scipostdeepblue} ~Submission }}

\fancyfoot[C]{\textbf{\thepage}}
}

\definecolor{tdo_blue}{HTML}{0000FF}
\definecolor{tdo_darkgreen}{HTML}{839A00}
\definecolor{tdo_red}{HTML}{FF0000}

\begin{document}

\pagestyle{SPstyle}

\begin{center}{\Large \textbf{\color{scipostdeepblue}{
Power-Law Spectra and Asymptotic \texorpdfstring{$\omega/T$}{omega/T} Scaling in the Orbital-Selective Mott Phase of a Three-Orbital Hubbard Model
}}}\end{center}
 
\begin{center}\textbf{
Fabian Eickhoff
}\end{center}

\begin{center}
Institute of Software Technology, German Aerospace Center, 51147 Cologne, Germany
\\[\baselineskip]
\href{mailto:email1}{\small fabian.eickhoff@dlr.de}
\end{center}

\section*{\color{scipostdeepblue}{Abstract}}
\boldmath\textbf{%
Quantum materials whose properties lie beyond the celebrated Landau Fermi-liquid paradigm have been observed for decades across diverse material platforms. 
Finding microscopic lattice models for metallic states that exhibit such peculiar behavior remains a major theoretical challenge, as these features often originate from strong quantum fluctuations in strongly interacting electron systems. 
Here we investigate a three-orbital Hubbard model at a high-symmetry point that hosts a transition from a metallic to an orbital-selective Mott (OSM) phase. Employing single-site dynamical mean-field theory combined with full-density-matrix numerical renormalization group, we chart the \(T\)–\(U\) phase diagram and obtain high-resolution real-frequency dynamics.
In the OSM regime we find asymptotically scale-invariant (power-law) single-particle spectra and asymptotic \(\omega/T\) scaling in both charge and spin channels, spanning several decades in frequency and temperature.
}

\vspace{\baselineskip}



\vspace{10pt}
\noindent\rule{\textwidth}{1pt}
\tableofcontents
\noindent\rule{\textwidth}{1pt}
\vspace{10pt}

\section{Introduction}
\label{sec:intro}

Strong electron–electron interactions govern the low–energy physics of a broad swath of quantum materials, including cuprate~\cite{Keimer2015} and iron–based superconductors~\cite{Analytis2014}, heavy–fermion compounds~\cite{Gegenwart2008}, transition–metal oxides~\cite{Imada1998}, and organic charge–transfer salts~\cite{Powell2011}. These systems host emergent phases—from unconventional superconductivity and putative spin liquids to metallic states that violate Landau’s Fermi–liquid expectations. Such anomalous metals frequently arise near Mott regimes and/or competing orders, where quantum criticality often provides a unifying framework for the observed power laws and scaling collapses in thermodynamic, transport, and dynamical probes.

A hallmark of scale invariance is the collapse of dynamical responses when plotted versus \(\omega/T\), indicating temperature as the dominant low–energy scale. In heavy–fermion metals, inelastic neutron scattering on CeCu\(_{6-x}\)Au\(_x\) revealed a dynamical spin susceptibility \(\chi''(\mathbf{Q},\omega,T)\) consistent with \(\omega/T\) scaling at an antiferromagnetic quantum critical point (QCP)~\cite{Schroeder2000}. In the cuprates, optical studies uncovered \(\omega/T\) scaling of the conductivity over wide \((\omega,T)\) windows in the strange–metal regime~\cite{vdMarel2003,Michon2023,Heumen2022}. Related magnetic scaling near putative QCPs appears in iron pnictides such as BaFe\(_2\)(As\(_{1-x}\)P\(_x\))\(_2\)~\cite{Hu2018_BaFe2AsP}. Additional examples include layered cobalt oxides, where a low–energy optical mode displays \(\omega/T\) scaling up to room temperature~\cite{Limelette2013}. Taken together, these observations show that \(\omega/T\) scaling is not confined to a single materials class or probe.

Theoretically, \(\omega/T\) scaling is natural in regimes dominated by quantum–critical fluctuations, where temperature provides the sole infrared cutoff. It is expected when hyperscaling holds near interacting fixed points and is compatible with local/boundary criticality, where imaginary–time correlators often assume conformal forms. At the impurity level, pseudogap Anderson and Kondo models host interacting critical points with \(\omega/T\) scaling in charge and spin dynamics~\cite{IngersentSi2003,Fritz2006,Glossop2011}. In lattice settings, beyond phenomenologies such as the marginal Fermi liquid~\cite{Varma1989_PRL}, controlled field–theoretic and (extended/cluster) DMFT treatments predict \(\omega/T\) scaling for spin and/or charge responses in quantum–critical fans above finite–\(T\) end points or \(T=0\) QCPs~\cite{Sachdev2011_QPT,Si2001_Nature,Gleis2024,Gleis2025}. For example, DMFT for the half–filled Hubbard model yields a finite–\(T\) Mott end point above which local spin and charge dynamics can exhibit \(\omega/T\) collapse~\cite{Dasari2017PRB}, consistent with approximate local quantum criticality~\cite{Eisenlohr2019}. Constructions inspired by Sachdev–Ye–Kitaev physics or holography provide complementary paradigms with explicit scale–invariant responses (often of \(\omega/T\) form), albeit typically relying on randomness or all–to–all couplings rather than microscopic short–range lattice physics~\cite{Song2017_PRL,Patel2018_PRX,Hartnoll2015_PRB}. Despite this breadth, clean microscopic lattice Hamiltonians with purely local interactions that produce robust \(\omega/T\) scaling—especially in the single–particle sector and over many decades in \(\omega/T\)—remain scarce. By contrast, Gaussian Hertz–Millis theory does not generically yield simple \(\omega/T\) scaling~\cite{Hertz1976,Millis1993}.

Motivated by these considerations, we study a three–orbital Hubbard model in which two symmetry–related conduction bands hybridize locally with a correlated \(f\) band. This symmetry stabilizes an orbital–selective Mott (OSM) phase—equivalently, a Kondo–breakdown phase in heavy–fermion language~\cite{Eickhoff2024_SciPost}. Using single–site DMFT combined with full–density–matrix NRG (FD–NRG), we show that the OSM phase exhibits power–law single–particle spectra at \(T=0\) and a striking asymptotic \(\omega/T\) collapse at finite \(T\), spanning several decades in \(\omega/T\). By “asymptotic” we mean that the scaling originates from a self–consistently generated pseudogap impurity problem in the local–moment regime rather than from proximity to a lattice QCP. Correspondingly, \(\omega/T\) scaling is realized for \(T\!\lesssim\! T^*\), with a crossover scale \(T^*\) that remains finite—but large—throughout the OSM phase.

\section{Model}
\label{sec:model}

Following Ref.~\cite{Eickhoff2024_SciPost}, we consider a minimal three–band
Hubbard Hamiltonian in which two symmetry-related, noninteracting
conduction bands couple locally to a single, strongly correlated \(f\) band:
\begin{align}
    \label{eq:H}
    H = H_\mathrm{bands} + H_\mathrm{hyb} + H_\mathrm{int}.
\end{align}
The three contributions read
\begin{align}
    H_\mathrm{bands} &= \sum_{k,\sigma} \epsilon_k \bigl(c^\dagger_{1,k,\sigma}c_{1,k,\sigma} - c^\dagger_{2,k,\sigma}c_{2,k,\sigma} + f^\dagger_{k,\sigma}f_{k,\sigma}\bigr),
    \label{eq:H_bands}\\
    H_\mathrm{hyb} &= \sum_{k,\sigma}\sum_{i\in\{1,2\}} \Bigl(V\, f^\dagger_{k,\sigma} c_{i,k,\sigma} + \mathrm{h.c.}\Bigr),\\
    H_\mathrm{int} &= U \sum_{l} n^f_{l\uparrow} n^f_{l\downarrow} - \epsilon_f \sum_{l,\sigma} n^f_{l\sigma}.
\end{align}
Here \(c_{1,k,\sigma}\) (\(c_{2,k,\sigma}\), \(f_{k,\sigma}\)) annihilates an electron with momentum \(k\) and spin \(\sigma\) in the respective band with dispersion \(\pm\epsilon_k\); the two noninteracting \(c\) bands hybridize on site with the \(f\) band via \(V\), and the \(f\) electrons experience a local repulsion \(U\).

The sign difference in Eq.~\eqref{eq:H_bands} enforces destructive interference in the real part of the hybridization, producing a symmetry-protected node at \(\omega=0\).
This node precludes Kondo screening and stabilizes an
orbital–selective Mott (OSM) phase without Hund’s coupling already within single-site DMFT~\cite{Eickhoff2024_SciPost}.
A simplified mechanism is outlined in Appendix~\ref{app:SimplePicture}.

While this band-structure symmetry may seem special, it naturally arises in systems 
with enlarged real-space unit cells, such as the depleted periodic Anderson model 
at half filling~\cite{Eickhoff2025_Moire}. 
In general, back-folding of the electronic band structure into a reduced Brillouin zone 
can generate analogous interference conditions. 
For a more detailed discussion, we refer to Sec.~4 of Ref.~\cite{Eickhoff2024_SciPost} 
and Appendix~A of Ref.~\cite{Eickhoff2025_Moire}.
Note that in contrast to Ref.~\cite{Eickhoff2024_SciPost}, we here include a finite \(f\)-band dispersion, rendering all three bands itinerant.

Despite the symmetry constraints, the model captures several materials-relevant limits, including flat-band physics and regimes supporting both antiferromagnetic (AF) and ferromagnetic (FM) correlations.
In the limit of a vanishing \(f\)-orbital bandwidth, the noninteracting spectrum hosts a strictly flat band; for concrete real-space realizations of this class, rigorous flat-band mechanisms predict an FM ground state~\cite{Potthoff1,Potthoff2}.
By contrast, a finite \(f\)-bandwidth naturally generates antiferromagnetic superexchange between correlated orbitals, favoring AF ground states within the OSM phase.
Strange, non–Fermi–liquid features are indeed reported in flat-band materials such as magic-angle twisted bilayer graphene~\cite{Cao2020,Jaoui2022}, in strange metals near antiferromagnetic domes such as the high-\(T_c\) cuprates~\cite{Mirarchi2022,Heumen2022,Phillips2022}, and also in ferromagnetic heavy-fermion compounds~\cite{Shen2020}.

We solve the strongly interacting lattice problem using single-site dynamical mean-field theory (DMFT)~\cite{Georges1996_RMP}, which maps the lattice Hamiltonian in Eq.~\eqref{eq:H} onto a self-consistent quantum impurity model and enables atomistic first-principles calculations for correlated materials~\cite{DMFT_FP_1, DMFT_FP_2, DMFT_FP_3, DMFT_FP_4, DMFT_FP_5, DMFT_FP_6}.
To solve the resulting impurity problem we employ the Numerical Renormalization Group (NRG), as implemented in the open-source NRG Ljubljana package developed by Rok Žitko and collaborators~\cite{NRG1, NRG2}. Details of the DMFT equations and their numerical implementation are provided in Apps.~\ref{subsec:method} and~\ref{app:numerics}.

\section{Results}
\label{sec:results}

Here we will present our results on the phase diagram and the finite $T$ dynamics.
These results will be further analyzed and discussed in Sec.~\ref{sec:discussion}.

Throughout this section we consider a Bethe lattice with infinite coordination number and measure energies in units of the half–bandwidth, \(D \equiv 1\). In line with the original analysis of the three–band OSM phase in Ref.~\cite{Eickhoff2024_SciPost}, the qualitative features discussed below are largely insensitive to the choice of lattice geometry and its corresponding density of states. We keep the hybridization fixed at \(V=D/4\) and drive the transition by increasing the on–site interaction \(U\).

\subsection{Phase diagram at particle–hole symmetry}
\label{subsec:phase_diagram}

\begin{figure}[t!]
    \centering
    \includegraphics[width=0.98\textwidth]{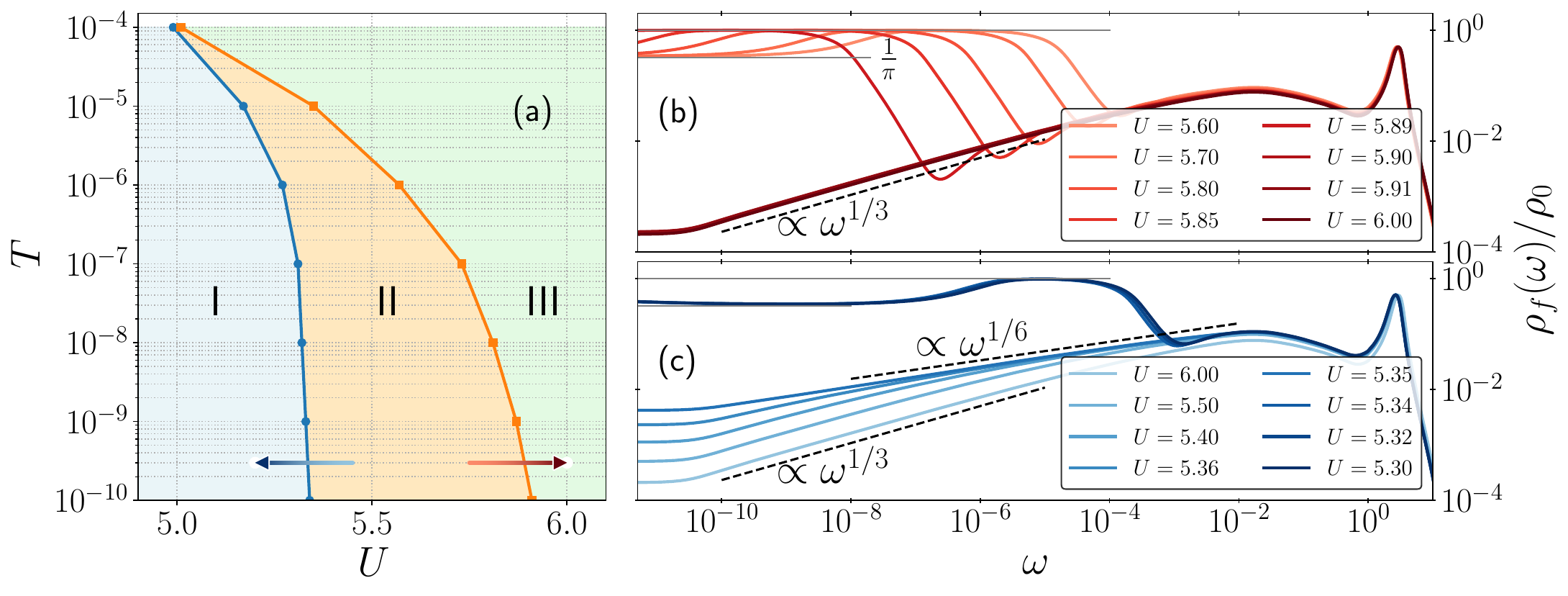}
    \caption{\textbf{Phase diagram and local \(f\)–orbital spectra across the orbital–selective Mott (OSM) transition.}
    (a) Phase diagram in the \(T\)–\(U\) plane showing regions I (metal), II (coexistence), and III (orbital selective insulator with power–law spectrum). 
    (b) \(f\)–orbital spectra at \(T=10^{-10}\) across the II\(\to\)III boundary, initialized from the metallic solution.
    (c) \(f\)–orbital spectra at \(T=10^{-10}\) across the II\(\to\)I boundary, initialized from the power–law solution.}
    \label{fig:1}
\end{figure}

Figure~\ref{fig:1}(a) shows the phase diagram in the \(T\)–\(U\) plane at particle hole symmetry, $\epsilon_f=-U/2$. Three regimes appear, labeled I, II, and III (shaded in different colors).  
In region~I (light blue), DMFT admits a single metallic solution in which the real–frequency \(f\)–orbital spectrum,
\(\rho_f(\omega) = -\tfrac{1}{\pi}\mathrm{Im}\,G^{R}_{f}(\omega)\), exhibits a quasiparticle resonance centered at \(\omega=0\). At low temperature this resonance is pinned to \(\rho_0/\pi\), where \(\rho_0\) denotes the noninteracting density of states of the Bethe lattice at the Fermi energy, \(\rho_0 = 2/(\pi D)\).  
In region~III (light green), a single solution is found whose \(f\)–orbital spectrum follows a power law \(\propto |\omega|^{1/3}\) down to \(|\omega|\!\sim\!T\), below which it crosses over to a constant; thus, at \(T=0\) this phase has insulating $f$-orbitals.  
Region~II (orange) is a coexistence regime: both the metallic and the power–law solutions can be stabilized, depending on the initial effective medium used in the DMFT self–consistency. The coexistence is bounded by two critical interactions, \(U_{c1}\) and \(U_{c2}\), and terminates at a finite–temperature critical endpoint, as expected for a first–order Mott transition within DMFT.

Figure~\ref{fig:1}(b) displays \(f\)–orbital spectra at \(T=10^{-10}\) across the II\(\to\)III boundary when the DMFT cycle is initialized in the metallic state. For \(U<U_{c2}\!\approx\!5.895\), the spectrum remains metallic with the zero–frequency value pinned to \(\rho_0/\pi\), while the resonance narrows upon approaching \(U_{c2}\). At the phase boundary, the resonance collapses and the spectrum crosses over to the \(|\omega|^{1/3}\) power law down to \(|\omega|\!\sim\!T\). In region~III (\(U>U_{c2}\)), the spectral shape depends only weakly on \(U\).

Figure~\ref{fig:1}(c) shows \(f\)–orbital spectra at \(T=10^{-10}\) across the II\(\to\)I boundary when the DMFT cycle is initialized in the power–law state. For \(U>U_{c1}\!\approx\!5.355\), the spectrum retains a power–law form; as \(U\to U_{c1}^{+}\), the apparent exponent drifts toward \(1/6\). - This is similar to the single–band Hubbard model, where the insulating solution posses a full gap in general, however, when $U_{c1}$ is approached from the insulating side a power–law spectrum emerges~\cite{Eisenlohr2019}. -
At the boundary, the spectral features change discontinuously; for \(U<U_{c1}\) the metallic solution becomes stable again and shows only a weak dependence on \(U\).

\subsection{\texorpdfstring{$\omega/T$}{omega/T} scaling in the orbital-selective Mott phase}

\subsubsection{particle-hole symmetric \texorpdfstring{$f$}{f}-orbitals}

\begin{figure}[t!]
    \centering 
    \includegraphics[width=0.98\textwidth]{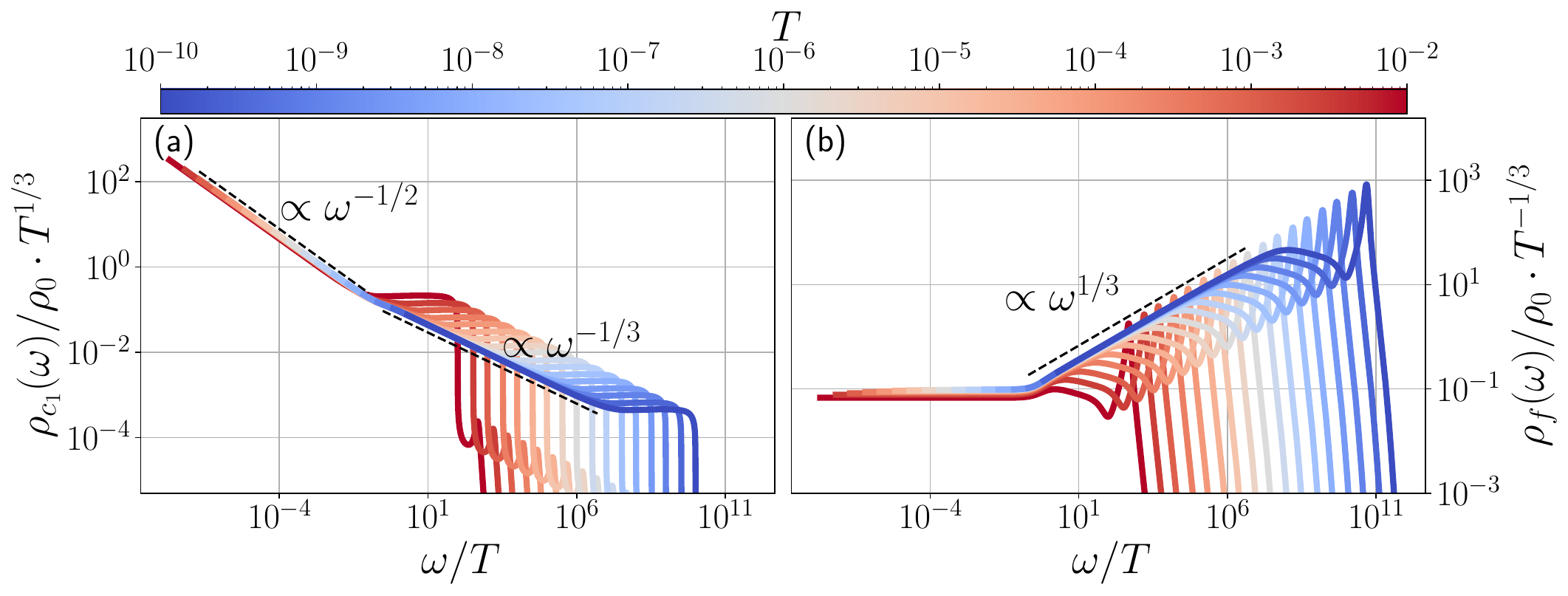}
    \caption{\textbf{\(\omega/T\) scaling of single–particle spectra in the OSM phase at particle–hole symmetry.}
    \textbf{(a)} \(c_1\)-orbital spectra plotted as \(\rho_{c_1}(\omega)/\rho_0 \cdot T^{1/3}\) versus \(\omega/T\) collapse onto a single curve over many decades. Guide lines highlight \(|\omega/T|^{-1/3}\) at large \(|\omega|/T\) and \(|\omega/T|^{-1/2}\) at small \(|\omega|/T\).
    \textbf{(b)} \(f\)-orbital spectra plotted as \(\rho_{f}(\omega)/\rho_0 \cdot T^{-1/3}\) versus \(\omega/T\) collapse onto a universal curve with \(|\omega/T|^{+1/3}\) at large \(|\omega|/T\) and a constant plateau at small \(|\omega|/T\), consistent with \(\rho_f(\omega=0,T)\sim T^{1/3}\). The interaction is fixed, $U=10>U_{c1}$, while colors encode temperature (blue: low \(T\); red: high \(T\)).}
    \label{fig:2}
\end{figure}

We now analyze dynamical scaling inside the OSM phase at particle–hole symmetry and fix the interaction to \(U=10>U_{c2}\).
As the spectra of both $c$-orbitals are qualitatively similar, we focus on the $c_1$-orbital only.
For a wide range of temperatures, the single–particle spectra collapse onto universal scaling functions of the ratio \(\omega/T\).
Specifically, the spectra depicted in Fig.~\ref{fig:2} are well described by
\begin{equation}
\rho_{c_1}(\omega,T)=T^{-1/3}\, F_{c_1}(\omega/T),\qquad
\rho_{f}(\omega,T)=T^{+1/3}\, F_{f}(\omega/T),
\end{equation}
so that \(T^{1/3}\rho_{c_1}\) and \(T^{-1/3}\rho_{f}\) become temperature–independent functions of \(\omega/T\). In our plots we normalize by \(\rho_0\), the Bethe–lattice density of states at the Fermi level, and display \(\rho_{c_1}(\omega)/\rho_0 \cdot T^{1/3}\) and \(\rho_{f}(\omega)/\rho_0 \cdot T^{-1/3}\) versus \(\omega/T\). This reveals a robust data collapse across many decades in \(\omega/T\) and over temperatures spanning roughly \(10^{-10}\) to \(10^{-3}\) (blue to red).

The asymptotics of the scaling functions follow clear power laws.
For \(|\omega|\gg T\), we find
\(\rho_{c_1}(\omega)\propto |\omega|^{-1/3}\) and \(\rho_f(\omega)\propto |\omega|^{+1/3}\).
For \(|\omega|\lesssim T\), the \(c_1\)-spectrum crosses over to a steeper \(|\omega|^{-1/2}\) behavior in the rescaled representation, while the \(f\)-spectrum approaches a \(T^{1/3}\)-controlled plateau, i.e., it is essentially \(\omega\)-independent after rescaling. These features are evident from the reference slopes indicated in panels (a) and (b).

In Appendix~\ref{app:NRG_check} we show the electron self-energy and verify that the
power-law behavior and $\omega/T$ scaling reported in the main text are insensitive to the
choice of NRG hyperparameters, confirming the numerical robustness of our results, aside
from small and expected shifts of the exact phase boundaries.

\begin{figure}[t!]
    \centering
    \includegraphics[width=0.98\textwidth]{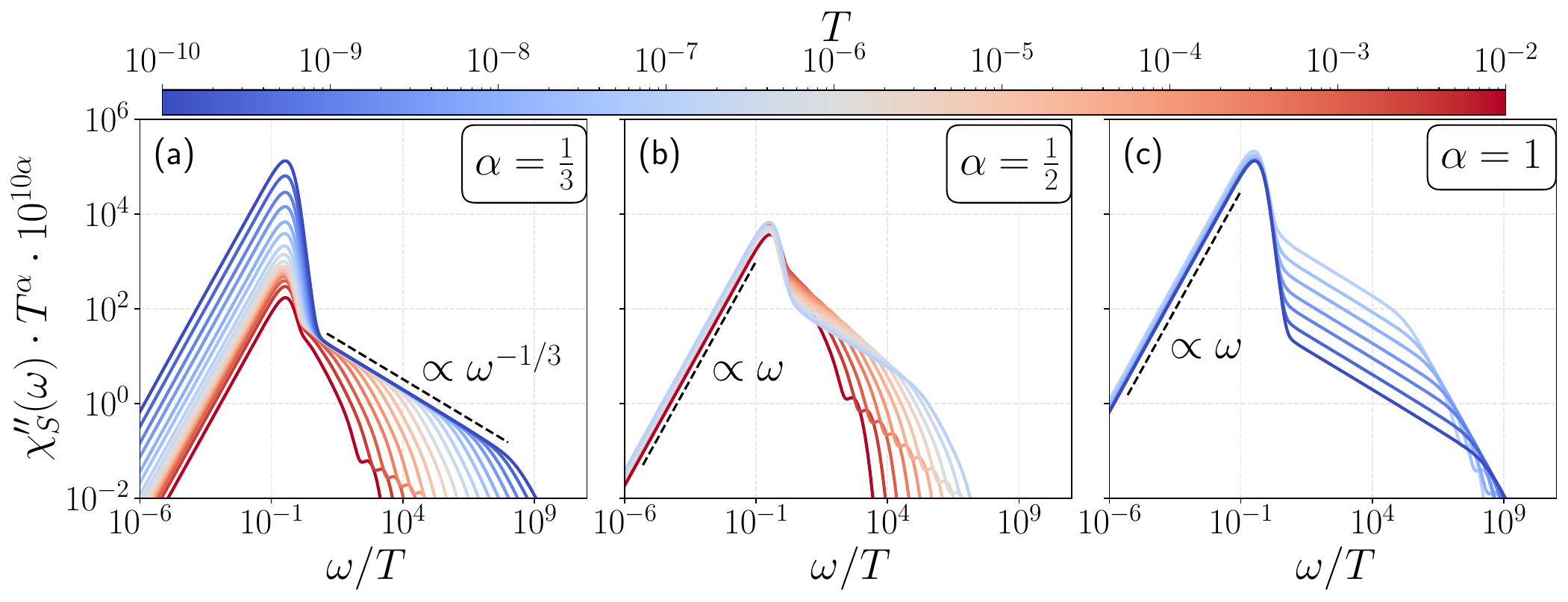}
    \caption{\textbf{Local dynamical spin susceptibility in the OSM phase at particle–hole symmetry (\(U=10>U_{c2}\)).}
    Temperature-dependent curves of the imaginary part of the local spin susceptibility, \(\chi''_{S}(\omega,T)\), are plotted against \(\omega/T\) after rescaling by \(T^{\alpha}\) (an additional factor \(10^{10\alpha}\) is applied for better comparison).
    \textbf{(a)} With \(\alpha=\tfrac{1}{3}\), the high-frequency tail collapses and follows \(\chi''_{S}\propto |\omega|^{-1/3}\) for \(|\omega|/T\gg1\).
    \textbf{(b)} With \(\alpha=\tfrac{1}{2}\), the low-frequency sector collapses for intermediate temperatures (\(T\geq 10^{-7}\)) and is linear, \(\chi''_{S}\propto \omega\), for \(|\omega|/T\lesssim1\).
    \textbf{(c)} With \(\alpha=1\), the low-frequency sector for ultra-low temperatures (\(T\leq 10^{-7}\)) again shows \(\chi''_{S}\propto \omega\), indicating a change in the temperature scaling exponent while preserving the linear \(\omega\) dependence.
    Colors encode temperature (blue: low \(T\); red: high \(T\)).}
    \label{fig:3}
\end{figure}

We next examine the local dynamical spin susceptibility in the OSM phase at particle–hole symmetry and \(U=10>U_{c2}\).
For each temperature we compute the imaginary part of the retarded local spin response, \(\chi''_{S}(\omega,T)\), and test a scaling form
\begin{equation}
\chi''_{S}(\omega,T) \;=\; T^{-\alpha}\,\Phi(\omega/T),
\end{equation}
by plotting \(T^{\alpha}\chi''_{S}\) versus \(\omega/T\) in Fig.\ref{fig:3}.
A data collapse is observed in three complementary regimes:

(i) \textit{High-frequency tail:} Using \(\alpha=\tfrac{1}{3}\) produces a collapse for \(|\omega|/T\gg1\) with a clear power law \(\chi''_{S}(\omega)\propto |\omega|^{-1/3}\).

(ii) \textit{Low-frequency, intermediate-\(T\) regime:} For \(T\geq 10^{-7}\), choosing \(\alpha=\tfrac{1}{2}\) collapses the curves for \(|\omega|/T\lesssim1\), revealing a linear low-frequency form, \(\chi''_{S}(\omega,T)\propto T^{-1/2}(\omega/T)\).

(iii) \textit{Low-frequency, ultra-low-\(T\) regime:} For \(T\leq 10^{-7}\) the optimal collapse at \(|\omega|/T\lesssim1\) is obtained with \(\alpha=1\), again linear in frequency, \(\chi''_{S}(\omega,T)\propto T^{-1}(\omega/T)\).

Thus the small-\(\omega\) sector remains linear in \(\omega\), while the temperature scaling dimension crosses over from \(\alpha=\tfrac{1}{2}\) to \(\alpha=1\) upon lowering \(T\).

\subsubsection{particle-hole asymmetric \texorpdfstring{$f$}{f}-orbitals}

\begin{figure}[]
    \centering
    \includegraphics[width=0.98\textwidth]{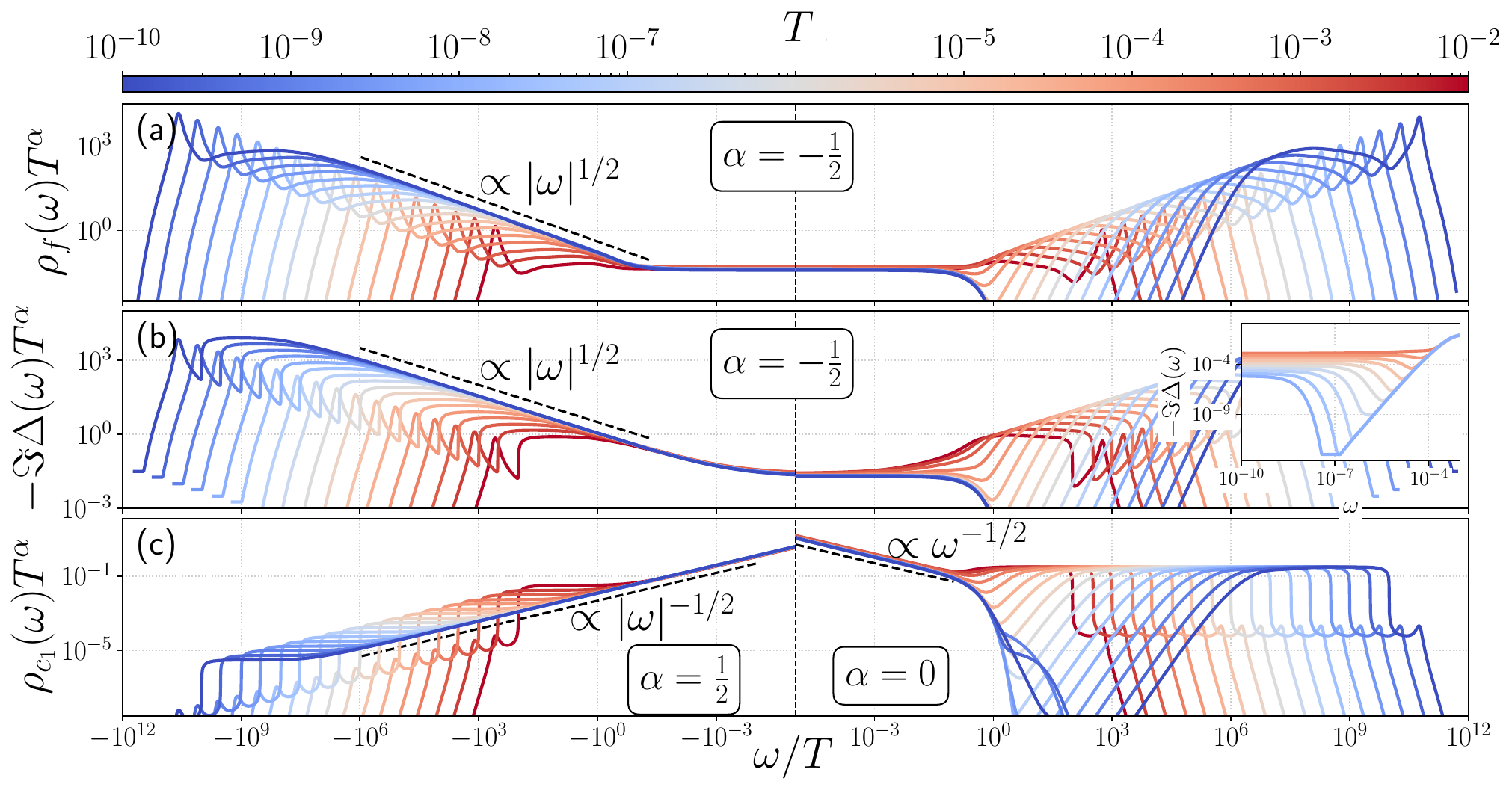}
    \caption{\textbf{\(\omega/T\) scaling at particle–hole asymmetry in the OSM phase.}
    Spectra are plotted on log–log axes using a mirrored abscissa: for \(\omega<0\) we show \(-\log|\omega/T|\) (hole side, left), and for \(\omega>0\) we show \(\log(\omega/T)\) (particle side, right).
    Curves are color–coded by temperature (blue: lowest \(T\); red: highest \(T\)).
    \textbf{(a)} \(f\)-orbital spectrum: \(T^{-1/2}\rho_f(\omega,T)\).
    \textbf{(b)} effective medium: \(T^{-1/2}\!\left[-\mathrm{Im}\,\Delta(\omega,T)\right]\).
    \textbf{(c)} \(c_1\)-orbital spectrum: hole–side collapse for \(T^{+1/2}\rho_{c_1}(\omega,T)\), and an additional particle–side low–frequency collapse for \(0<\omega\lesssim T\) when plotted without a temperature prefactor (effective exponent \(\alpha=0\)).}
    \label{fig:4}
\end{figure}

We now break particle–hole (PH) symmetry while remaining in the OSM phase by fixing \(U=10\) and shifting the \(f\)-level to \(\varepsilon_f=-4\).
Frequencies are measured relative to the chemical potential such that \(\omega<0\) denotes the occupied (hole excitation) side and \(\omega>0\) the unoccupied (particle excitation) side.
All spectra in Fig.~\ref{fig:4} are shown on log–log axes; to display both signs on a logarithmic scale we use a mirrored representation, plotting the abscissa as \(-\log|\omega/T|\) for \(\omega<0\) and \(\log(\omega/T)\) for \(\omega>0\) (i.e., \(|\omega|/T\) on a log scale, mirrored about \(\omega=0\)).

In this PH-asymmetric case, we find robust \(\omega/T\) scaling on the hole side for all three dynamic quantities considered:
the \(f\)-orbital spectrum, the effective medium at self consistency, and the \(c_1\)-orbital spectrum.
Specifically, for \(\omega<0\) plotting \(T^{-1/2}\rho_f\), \(T^{-1/2}[-\mathrm{Im}\,\Delta]\), and \(T^{+1/2}\rho_{c_1}\) versus $|\omega|/T$ results in a nice data collapse for several decades of temperature (left halves of the panels). The high–frequency tails of the $f$-spectra obey the common asymptotic \(|\omega/T|^{1/2}\), while for \(|\omega|/T\lesssim1\) the rescaled curves approach constants, consistent with thermal cutoffs of the scaling functions.
The hole spectra of $c_1$-orbital is temperature independent, and,  thus $\propto \omega^{-1/2}$ down to $\omega=0$.

On the particle side, scaling is more selective.
While we still find \(\rho_f(\omega<T,T)\propto \sqrt{T}\) and \(-\mathrm{Im}\,\Delta_f(\omega<T,T)\propto\sqrt{T}\), the $c_1$-orbitals still show power law scaling for $\omega<T$ but with a different exponent $\alpha=0$: $\rho_{c_1}(\omega<T,T)\propto \sqrt{T/\omega}$.

At Frequencies $\omega>T$ all three spectra feature a sharp gap, as depicted in the inset of panel (b) for the imaginary part of the effective medium at self consistency. The relative width scales as $T$, while the total width is determined by the degree of particle-hole asymmetry (not shown).

\section{Discussion}
\label{sec:discussion}  

The results presented in Sec.~\ref{sec:results} demonstrate that the single-particle spectra and the effective medium in the OSM phase of the Hamiltonian in Eq.~\eqref{eq:H} exhibit a robust data collapse when plotted as $\omega/T$ and rescaled appropriately. In contrast, the local dynamical spin susceptibility shows $\omega/T$ scaling only at frequencies $\omega > T$, while the low-energy sector $\omega < T$ requires distinct rescaling exponents for $T > \tilde{T}^*$ and $T < \tilde{T}^*$, with $\tilde{T}^*$ denoting an additional (possibly very small) crossover scale.  

Within single-site DMFT, $\omega/T$ scaling can in principle arise in two ways. Since the $f$-orbital self-energy is determined self-consistently from the effective medium,  
\begin{align}
\Sigma(\omega,T) = \Sigma\!\big[\Delta_\text{eff}(\omega,T)\big],
\end{align}
the scaling could either originate from the temperature dependence of the medium itself or from intrinsic properties of the effective impurity problem. We have verified that in the present case at PH symmetry the latter mechanism applies: the finite-$T$ solution can be obtained from the effective medium at zero temperature ($T = 10^{-10}$ in the calculations),  
\begin{align}
\Sigma(\omega,T)\!\big[\Delta_\text{eff}(\omega,T)\big] 
\;\approx\; \Sigma(\omega,T)\!\big[\Delta_\text{eff}(\omega,0)\big].
\end{align}
Consequently, the finite-$T$ dynamics of the $f$-orbital spectra and the self-energy emerge entirely from the effective pseudo-gap impurity problem characterized by $\Delta_\text{eff}(\omega,0)$.  

\paragraph{Pseudo-gap SIAM.}  
The phase diagram of the single-impurity Anderson model (SIAM) with a pseudo-gap density of states characterized by an exponent $r<0.5$ is well established. It hosts an interacting non-Fermi-liquid fixed point with pronounced local-moment (LM) fluctuations~\cite{PseudoGapAnderson_1,PseudoGapAnderson_2,PseudoGapAnderson_3,PseudoGapAnderson_4}, and displays $\omega/T$ scaling in both the charge and spin sectors~\cite{PseudoGapAnderson_2}, at a critical interaction $U_c^\text{PSIAM}$. For $U<U_c^\text{PSIAM}$ the impurity spin is screened, while for $U>U_c^\text{PSIAM}$ it remains unscreened.  

In the OSM phase discussed here, however, DMFT self-consistency always stabilizes the LM phase (see Sec.~4.3.2 of Ref.~\cite{Eickhoff2024_SciPost}), raising questions about the origin of the observed scaling behavior.  

From the hyperscaling properties of the interacting fixed point in the pseudo-gap SIAM, the general scaling form of the single-particle spectral function is known~\cite{Fritz2006,Glossop2005}:  
\begin{align}
    \label{eq:rg_scaling}
    \rho_f(\omega) \propto \omega^{1-\eta}\,\Theta\!\left(\frac{\omega}{T},\frac{T}{T^*}\right),
\end{align}
where $\eta$ is a scaling dimension and $T^*$ denotes the crossover scale that vanishes at the critical point $U=U_c^\text{PSIAM}$.  

True quantum-critical $\omega/T$ scaling across all temperatures occurs only at $T^*=0$, i.e., precisely at the critical point. Nevertheless, Eq.~\eqref{eq:rg_scaling} also allows for asymptotic $\omega/T$ scaling in the regime $T/T^*\ll 1$, provided the scaling function admits an expansion in small parameter $T/T^*$:
\begin{align}
    \Theta \approx \phi_0(\omega/T) + (T/T^*)\phi_1(\omega,T) + \mathcal{O}\left( (T/T^*)^2\right)
\end{align}

\begin{figure}[htbp]
    \centering
    \includegraphics[width=1\textwidth]{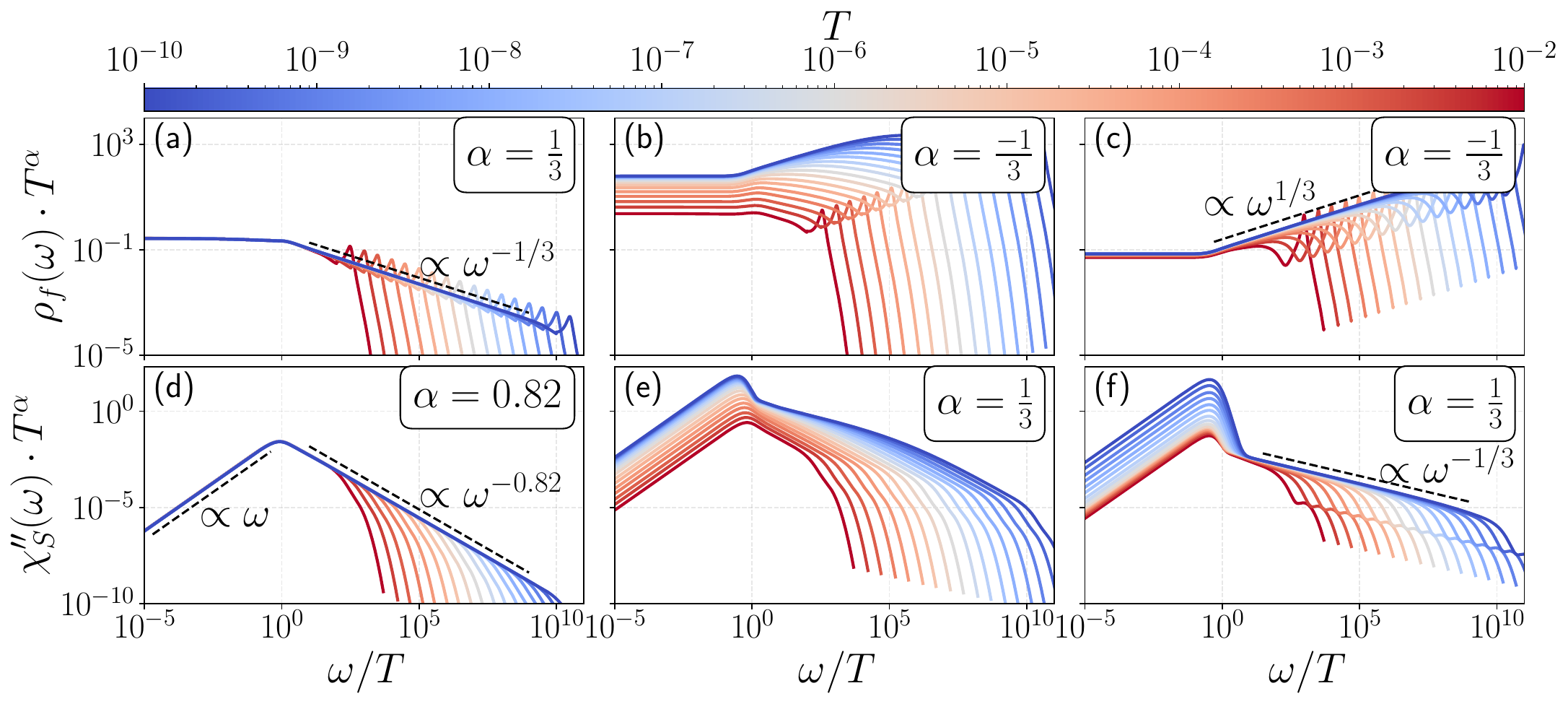}
    \caption{\textbf{Single-particle spectra and dynamic spin susceptibility of the pseudo-gap SIAM.}  
    Results are shown for the spectral function (a–c) and the imaginary part of the local spin susceptibility (d–f). Panels (a,d) display data at the quantum critical point $U=U_c^\text{PSIAM}\approx 5.784$, panels (b,e) close to the QCP on the LM side with $U=7\gtrsim U_c^\text{PSIAM}$, and panels (c,f) deep in the LM phase with $U=20\gg U_c^\text{PSIAM}$. All spectra are plotted as functions of $\omega/T$ and rescaled by $T^\alpha$ with exponents $\alpha$ chosen for optimal data collapse.}
    \label{fig:5}
\end{figure}
 
As far as we are aware, accurate numerical results on the scaling properties of the LM phase of the pseudo-gap SIAM are scarce. We therefore revisit the pseudo-gap SIAM defined by the hybridization function  
\begin{align}
    -\Im\Delta_{\text{imp}}(\omega) = \begin{cases}
    \tfrac{1}{3}|\omega/D|^{1/3}, & \text{if } |\omega| \leq D \\
    0 & \text{else}
\end{cases}
\end{align}
which yields $U_c^\text{PSIAM}\approx 5.784$. Figure~\ref{fig:5} summarizes our results.  

At the QCP [panels (a,d)], perfect $\omega/T$ scaling is observed when rescaling by $T^\alpha$ with $\alpha=1/3$ for $\rho_f$ and $\alpha=0.82$ for $\chi''$. In agreement with Ref.~\cite{Glossop2011}, the scaling function tends to a finite constant for $\rho_f$ and vanishes for $\chi''$ as $\omega/T \to 0$. Notably, the finite $\rho_f(0,T)$ at criticality contrasts with perturbative expectations~\cite{Glossop2005} and Callan–Symanzik resummation results~\cite{Fritz2006}.  

On the LM side close to the QCP ($U=7$, panels (b,e)), we do not obtain full data collapse across all $T$, but at very low $T\lesssim 10^{-6}$ the spectra converge toward a scaling function with exponent $\alpha=-1/3$. Deep in the LM phase ($U=20$, panels (c,f)), the finite-$T$ spectra for $\rho_f$ and $\chi''$ resemble those in the OSM phase [cf. Fig.~\ref{fig:2}(b) and Fig.~\ref{fig:3}(a)]. Here the single-particle spectra collapse over the full temperature range, while $\chi''$ exhibits scaling only over restricted frequency intervals.  

We conclude that the LM phase of the pseudo-gap SIAM admits asymptotically scale-invariant single-particle spectra with $\omega/T$ scaling, while in the spin response it emerges only in the limit $T/T^*\to 0$ within finite frequency windows. The crossover scale $T^*$ increases with distance from the QCP; deep in the LM phase, $T^*$ can become large enough that $\omega/T$ scaling appears already at intermediate temperatures.  

\begin{figure}[htbp]
    \centering
    \includegraphics[width=1\textwidth]{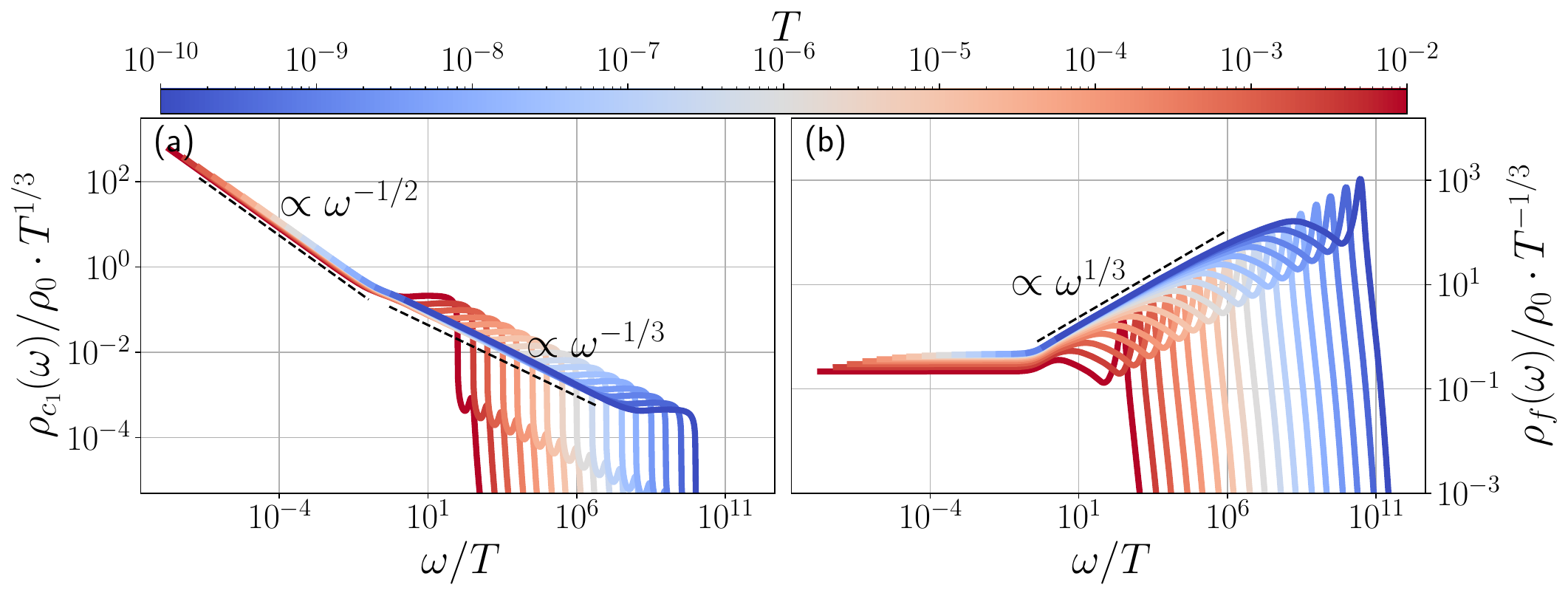}
    \caption{\textbf{\(\omega/T\) scaling of single–particle spectra in the OSM phase near the phase boundary.}  
    Same as Fig.~\ref{fig:2}, but for $U=6$ close to the OSM phase boundary at $U_{c2}\approx 5.9$.}
    \label{fig:6}
\end{figure}

Therefore, the $\omega/T$ scaling observed in the OSM phase of Eq.~\eqref{eq:H} does not reflect genuine quantum-critical behavior but rather asymptotic scaling. Unlike in the pseudo-gap SIAM, however, the crossover scale $T^*$ in the OSM phase is not solely fixed by the distance to the phase boundary: here $T^*$ is determined self-consistently through the effective medium. Figure~\ref{fig:6} illustrates this point: for $U=6$ near $U_{c2}$ the scaling closely resembles that deep in the OSM phase ($U=10$, Fig.~\ref{fig:2}), with only moderate deviations at the highest $T$. This indicates a mild increase of the crossover scale, $T^*(U=6)\lesssim T^*(U=10)$.  

In Ref.~\cite{Eickhoff2024_SciPost} we also analyzed the case of small deviations from 
perfectly symmetric hybridization, $V_1^2/D_1 \neq V_2^2/D_2$, where $V_i$ and $D_i$ denote 
the hybridization strengths and bandwidths of the two $c_i$ orbitals.  
Such asymmetry removes the exact cancellation of the real part of the hybridization 
function and therefore eliminates the OSM phase strictly at $T=0$.  
However, the associated low-energy crossover scale below which this breakdown occurs is 
exponentially small, so that the OSM phase remains effectively intact over a broad 
intermediate-energy window.

A further important scenario is the breaking of the exact inversion symmetry of the 
conduction bands, $\epsilon_k^1 \neq -\epsilon_k^2$.  
This modification, however, would prevent the reduction of the DMFT lattice integrals 
to one-dimensional forms, as used in Appendix~\ref{subsec:method}, and is therefore 
beyond the scope of the present work.  
Nevertheless, based on the simplified mechanism discussed in 
Appendix~\ref{app:SimplePicture}, we expect the OSM phase to remain stable at $T=0$ 
as long as inversion symmetry is restored in the vicinity of the Fermi surface.

\section{Summary and outlook}

Using single-site DMFT in combination with state-of-the-art FD-NRG, we have calculated the phase diagram of the three-band Hubbard model at a high-symmetry point in the \(T\)–\(U\) plane.  
We identified a coexistence regime separating a metallic from an orbital-selective Mott (OSM) phase.  
Within the OSM phase the zero-temperature single-particle spectra are asymptotically scale invariant, exhibiting power-law behavior, $\rho(\omega,T=0)\propto |\omega|^{r}$.  
The exponent $r$ differs between orbitals: while the $f$-orbital is insulating with $r=1/3$, the two conduction orbitals remain metallic with $r=-1/3$, highlighting the orbital-selective character of the Mott localization.  

We have further analyzed the finite-temperature dynamics inside the OSM phase with respect to possible $\omega/T$ scaling, for both particle–hole symmetric and asymmetric $f$-orbitals.  
Analogous to the local-moment (LM) phase of the pseudo-gap SIAM, we find asymptotic $\omega/T$ scaling in the single-particle spectral functions,  
\begin{align}
    T^\alpha \rho_i(\omega,T)\propto \phi_0(\omega/T) + (T/T^*)\phi_1(\omega,T) + \mathcal{O}\left( (T/T^*)^2\right)
\end{align}
with a crossover scale $T^*$.  
In contrast to the pseudo-gap SIAM, however, $T^*$ does not vanish at the OSM phase boundary but remains finite and relatively large throughout the OSM phase.  
As a result, $\omega/T$ scaling manifests already at moderate temperatures, rather than only asymptotically close to zero temperature.  

The behavior of the local dynamical spin susceptibility in the OSM phase is more intricate.  
While $T^\alpha \chi''(\omega,T)$ exhibits asymptotic $\omega/T$ scaling for $T/T^*\ll 1$ over intermediate frequency ranges, the low-energy sector requires different temperature exponents $\alpha=0.5$ and $\alpha=1$ depending on whether $T$ is larger or smaller than some energy scale $\tilde{T}^* < T^*$.

Our results demonstrate the realization of robust power-law spectra and asymptotic $\omega/T$ scaling in a microscopic lattice model with purely local interactions.  
Such scaling behavior is expected to strongly influence transport, potentially giving rise to $\omega/T$ scaling in the optical conductivity and non-Fermi-liquid dc resistivity.  
While our model relies on strong symmetry constraints in the band structure, it naturally includes a flat-band limit, and both ferromagnetic and antiferromagnetic phases are expected to emerge beyond the single-site DMFT approximation when band width of the localized orbital is tuned.  
This connection suggests a broader relevance: strange non-Fermi-liquid features in transport are indeed reported in flat-band materials such as magic-angle twisted bilayer graphene \cite{Cao2020, Jaoui2022}, in strange metals near antiferromagnetic domes such as the high-$T_c$ cuprates \cite{Mirarchi2022,Heumen2022,Phillips2022}, and in ferromagnetic heavy-fermion compounds \cite{Shen2020}.  
Numerical calculation of transport properties and future extensions incorporating cluster DMFT or non-local correlations could therefore provide important insights into these experimentally observed non-Fermi-liquid regimes.

\section*{Acknowledgements}
I thank the DLR–HPC–QCA group for constructive feedback. Special thanks go to Marius Stürmer and  Gonzalo Camacho for their support with the numerical calculations and for pointing me to the mpmath library.

\paragraph{Funding information}
This project was made possible by the Quantum Computing Initiative of the German Aerospace Center (DLR) and the Federal Ministry for Economic Affairs and Climate Action; \href{https://qci.dlr.de/alqu/}{qci.dlr.de/projects/ALQU}.

\paragraph{Data availability}
The data supporting the findings of this study are available from Zenodo at \cite{eickhoff_2025_17214657}
and also upon request from the author.

\begin{appendix}
\numberwithin{equation}{section}

\section{Dynamical mean-field treatment}
\label{subsec:method}

To capture local quantum fluctuations non-perturbatively we employ single-site
dynamical mean-field theory (DMFT) \cite{Georges1996_RMP}, which maps the lattice
problem \eqref{eq:H} onto a self-consistent single-impurity Anderson model (SIAM).

Starting from the matrix representation of the full lattice Green's function \( G(E, z) \) along the energy shell
\( E = \epsilon_k \): 
\begin{align}
\label{eq:G_matrix}
    G(E, z) = 
    \begin{pmatrix}
        z - E & 0 & V \\
        0 & z + E & V \\
        V & V & z - E - \Sigma(z)
    \end{pmatrix}^{-1}.
\end{align}
The local $f$ Green function $\mathcal{G}^{f}_\text{loc}(z)$ of the lattice model can be obtained from integrating the $f$-orbital contribution $G_{ff}(E,z)$:
\begin{equation}
  \mathcal{G}^{f}_\text{loc}(z)
  \;=\;
  \int_{-\infty}^{\infty}\!dE\,
     \rho_{0}(E)\,G_{ff}(E,z),
  \label{eq:G_loc}
\end{equation}
which is required to coincide with the impurity Green function
\begin{equation}
  \mathcal{G}^{f}_{\mathrm{imp}}(z) \;=\;
    \Bigl[
      z \!-\! \varepsilon_{f}
      - \Delta_{\mathrm{eff}}(z)
      - \Sigma(z)
    \Bigr]^{\!-1},
  \label{eq:G_imp}
\end{equation}
where $\Sigma(z)$ is the local electron self-energy and $\Delta_{\mathrm{eff}}(z)$ describes the effective medium.
Self-consistency demands $\mathcal{G}^{f}_{\mathrm{loc}}(z) = \mathcal{G}^{f}_{\mathrm{imp}}(z)$; thus
\begin{equation}
  \Delta_{\mathrm{eff}}(z)
  \;=\;
  z - \varepsilon_{f} - \Sigma(z)
  - \bigl[\mathcal{G}^{f}_{\mathrm{loc}}(z)\bigr]^{-1}.
  \label{eq:DMFT_SC}
\end{equation}

Starting from an initial $\Delta_{\mathrm{eff}}(z)$ we iterate:
  (I) solve the impurity problem to obtain $\Sigma(z)$ and $\mathcal{G}^{f}_{\mathrm{imp}}(z)$;
  (II) compute $\mathcal{G}^{f}_{\mathrm{loc}}(z)$ from Eq.~\eqref{eq:G_loc};
  (III) update $\Delta_{\mathrm{eff}}(z)$ via Eq.~\eqref{eq:DMFT_SC};
  (IV) repeat until convergence is reached.

\noindent
For details on the efficient and accurate computation of Eq.\eqref{eq:G_loc}, see Appendix~\ref{app:gff}.

To solve the effective SIAM, characterized by $\Delta_{\mathrm{eff}}(z)$ in Eq.\eqref{eq:DMFT_SC}, we employ the Numerical Renormalization Group (NRG) technique, as implemented in the open-source NRG Ljubljana code developed by Rok Žitko and collaborators\cite{NRG1, NRG2}.
In our NRG calculations, we exploited total spin and particle number conservation, combined with an improved discretization scheme~\cite{NRG2,NRG_discretization}.
The discretization parameter was set to $\lambda=2.5$, and 2000 states were retained at each NRG iteration.

To compute spectral functions, we employed the full-density-matrix NRG (FD-NRG) algorithm~\cite{NRG_Spectra1}, which makes use of a complete basis set of the Wilson chain~\cite{NRG_Spectra2}. Spectral features were broadened using the method described in Ref.~\cite{NRG_Spectra1}, with a broadening parameter $\alpha=0.8$.
Additionally, we apply the recently developed self-energy trick~\cite{NRG_selfEnergy}, which significantly improves numerical accuracy and enables state-of-the-art NRG spectral calculations.

\section{Numerical evaluation of lattice integrals}
\label{app:numerics}

All impurity–lattice self–consistency steps can be reduced to one–dimensional energy integrals.
In this appendix we show how these integrals
are evaluated numerically by means of a partial–fraction decomposition of the
three–band Green function.  Throughout we write complex frequencies as
$z=\omega+i\delta$, with $\delta=\mathcal{O}(10^{-15})$ smaller than any energy scale of interest, and suppress the (trivial) spin index.

\subsection{Partial–fraction form of the lattice Green function}
\label{app:pf}

Performing the inversion in Eq.~\eqref{eq:G_matrix} analytically can be rewritten in a compact form:
\begin{equation}
  \mathbf{G}_{ij}(E, z) = 
  \frac{\mathbf{Q}_{ij}(E, z)}{P(E, z)}, \qquad
  P(E, z) = \sum_{n=0}^{3} p_n(z)\, E^n,
  \label{eq:app_pf}
\end{equation}
where $P(E,z)$ is a cubic polynomial in~$E$ with
\begin{align}
    p_0(z) &= z^2(z-\Sigma(z)-\epsilon_f-2V^2),\\
    p_1(z) &= -z^2,\\
    p_2(z) &= \Sigma(z)-\epsilon_f-z,\\
    p_3(z) &= 1,
\end{align}
and the matrix numerator
$\mathbf Q(E,z)$ is at most quadratic. The two components relevant for the results presented in the main text read:
\begin{align}
    Q_{f,f}(E,z) &=(z-E)(z+E),\\
    Q_{c_1,c_1}(E,z) &= (z+E)(z-\epsilon_f-E-\Sigma(z))-V^2.
\end{align}

Let $\{\xi_{\ell}(z)\}$ (with $\ell=1,2,3$) be the complex roots of $P(E,z)$.
Because the roots are pairwise distinct for generic parameters one can write
\begin{equation}
  G_{ij}(E,z)\;=\;
  \sum_{\ell=1}^{3}\frac{R^{(ij)}_{\ell}(z)}{E-\xi_{\ell}(z)},
  \qquad
  R^{(ij)}_{\ell}(z)=
  \frac{Q_{ij}(\xi_{\ell},z)}{P'(\xi_{\ell},z)},
  \label{eq:app_res}
\end{equation}
with $P'$ the derivative with respect to~$E$.  The residues
$R^{(ij)}_{\ell}$ depend smoothly on the external frequency~$z$ and on the
self–energy $\Sigma(z)$ that enters through the $f$–orbital diagonal element of
$\mathbf H_{k}$.

The complex roots $\xi_{\ell}$ were obtained using Cardano’s formula.
For small frequencies $z=\omega+i\delta$, the coefficients $p_0(z)$ and $p_1(z)$
scale as $z^{2}$, reaching extremely small values for $\delta=10^{-15}$ and 
$\omega \rightarrow 0$.
This renders the cubic root problem numerically ill-conditioned in double precision.
To reliably resolve the two roots that collapse to zero in this limit, we evaluate 
Cardano’s formula with high-precision arithmetic using the multiprecision Python 
library \texttt{mpmath}~\cite{mpmath}.

\subsection{Local lattice Green function in the DMFT loop}
\label{app:gff}

In order to accurately and efficiently compute the local Greens function $\mathcal{G}^f_\text{loc}(z)$ from evaluating Eq.~\eqref{eq:G_loc} we substitute the partial fraction from of $G_{ff}(E,z)$ from Eq.\eqref{eq:app_res} to obtain
\begin{equation}
  \mathcal{G}^{f}_\text{loc}(z)
  \;=\;
  \sum_{\ell=1}^{3}R^{(ff)}_{\ell}(z)\,
   \mathcal H\bigl[\rho_{0}\bigr]\!\bigl(\xi_{\ell}(z)\bigr),
  \label{eq:app_gfloc}
\end{equation}
where $\mathcal H$ denotes the complex Hilbert transform
\[
  \mathcal H\bigl[\rho_{0}\bigr](\xi)
  =\int_{-\infty}^{\infty}\!dE\,
     \frac{\rho_{0}(E)}{\xi-E}.
\]
For the Bethe lattice with half–bandwidth $D=1$
one has
$\rho_{0}(E)=\tfrac{2}{\pi}\sqrt{1-E^{2}}$ and
\begin{equation}
  \mathcal H\bigl[\rho_{0}\bigr](\xi)
  =2\,\bigl(\xi-\sqrt{\xi^{2}-1}\bigr),
  \qquad
  \mathcal H^{\phantom{\dagger}}\!(\xi^{*})=
  \mathcal H(\xi)^{*},
  \label{eq:app_ht_bethe}
\end{equation}
so that Eq.\,\eqref{eq:app_gfloc} is obtained in closed form once the three
roots $\{\xi_{\ell}\}$ are known.  Equation~\eqref{eq:app_gfloc} replaces
numerical quadrature by simple residue algebra and is evaluated at each step
of the self–consistency cycle.

\vspace{0.5em}
\noindent
In summary, by decomposing every lattice Green function into simple poles and
using the analytic Hilbert transform of the Bethe density of states, all
energy integrals required for the DMFT self-consistency reduce to algebraic expressions involving only the roots $\xi_{\ell}$ and the residues $R^{(ab)}_{\ell}$.

\section{Simplified mechanism for restoring the OSM phase in multi-band models}
\label{app:SimplePicture}

Here we provide a simplified physical picture to explain why the three-orbital model is able to restore the orbital-selective Mott (OSM) phase within single-site DMFT. The following discussion is based on a series of approximations and is intended to offer a qualitative understanding rather than a quantitatively accurate description.

\paragraph{Critical interaction from linearized DMFT}

From linearized DMFT~\cite{Bulla2000}, one can obtain an analytical estimate of the critical interaction strength \( U_c \) for the Mott transition in a single-band model with hopping elements \( t_{ij} \) and DOS \( \rho(\epsilon) \):
\begin{align}
    \label{app:Uc}
    U_c = 6 \sqrt{\sum_j t_{ij}^2} = 6 \sqrt{\int \epsilon^2 \rho(\epsilon) \, \mathrm{d}\epsilon}.
\end{align}
This estimate has been shown to agree well with full DMFT calculations and highlights the importance of the high-energy tail of the DOS in obtaining a finite \( U_c < \infty \).

\paragraph{Effect of hybridization on the DOS}

Now consider the case where this band hybridizes with another band via a coupling \( V \). This hybridization effectively renormalizes both the hopping amplitudes and the dispersion:
electrons from one band can tunnel into the other, propagate there, and return, leading to modified effective hopping terms \( t_{ij} \to t'_{ij} \) and dispersion \( \epsilon_k \to \epsilon_k' \).

In the simplest approximation—neglecting retardation effects—this additional tunneling amplitude from site \( i \) to \( j \) can be estimated as \( V^2 \Re G_{ij}(\omega = 0) \), where \( G_{ij}(\omega) \) is the single-particle Green's function of the additional band. This results in the renormalized dispersion:
\[
\epsilon_k' = \epsilon_k + \frac{V^2}{\gamma_k},
\]
where \( \gamma_k \) is the dispersion of the additional band.

The corresponding renormalized DOS becomes:
\begin{align}
    \label{app:rho_p}
    \rho'(\epsilon) =\frac{1}{N} \sum_k \delta\left(\epsilon - \epsilon_k - \frac{V^2}{\gamma_k} \right).
\end{align}

\paragraph{Breakdown of the Mott transition}

At high energies \( \epsilon \gg D \) the \( \epsilon_k \) term becomes negligible and the effective DOS $\rho^\prime(\epsilon)$ is dominated by contributions from \( V^2/\gamma_k \). Using the identity \( \delta[g(x)] = \delta(x - x_0)/|g'(x_0)| \), we find:
\begin{align}
    \rho'(\epsilon \gg D) = \frac{V^2}{\epsilon^2} \rho_0\left(\frac{V^2}{\epsilon} \right).
\end{align}

Inserting this result into Eq.~\eqref{app:Uc} shows that the integral diverges as long as $\rho_0(0)$ is finite, implying:
\[
U_c = \infty,
\]
i.e., the Mott transition is suppressed due to the enhancement of the DOS at high energies caused by the hybridization.

\paragraph{Restoring a finite critical interaction}
To restore a finite \( U_c \), two strategies are possible:

\begin{enumerate}
    \item \textbf{Nodal Hybridization:} Introduce a hybridization function that vanishes at the Fermi level, e.g., \( V_k \propto \gamma_k \). This approach was analyzed in Ref.~\cite{Held2000}.

    \item \textbf{Destructive Interference:} Add a third band with dispersion \( \eta_k \), leading to a interference of the hybridization effects. The renormalized dispersion becomes:
    \[
    \epsilon_k' = \epsilon_k + \frac{V^2}{\gamma_k} + \frac{V^2}{\eta_k}.
    \]
    In certain cases, i.e. $\eta_k=-\gamma_k$, this effectively restores a finite support of $\rho^\prime(\epsilon)$ and thus a finite \( U_c \) when Eq.~\eqref{app:Uc} is evaluated again.
\end{enumerate}

This explains qualitatively how the presence of a third orbital can stabilize the OSM phase within single-site DMFT by preventing the unbounded growth of effective tunneling amplitudes through interference.

\section{Electron self-energy and NRG hyperparameter scan}
\label{app:NRG_check}

\begin{figure}[htbp]
    \centering
    \includegraphics[width=1\textwidth]{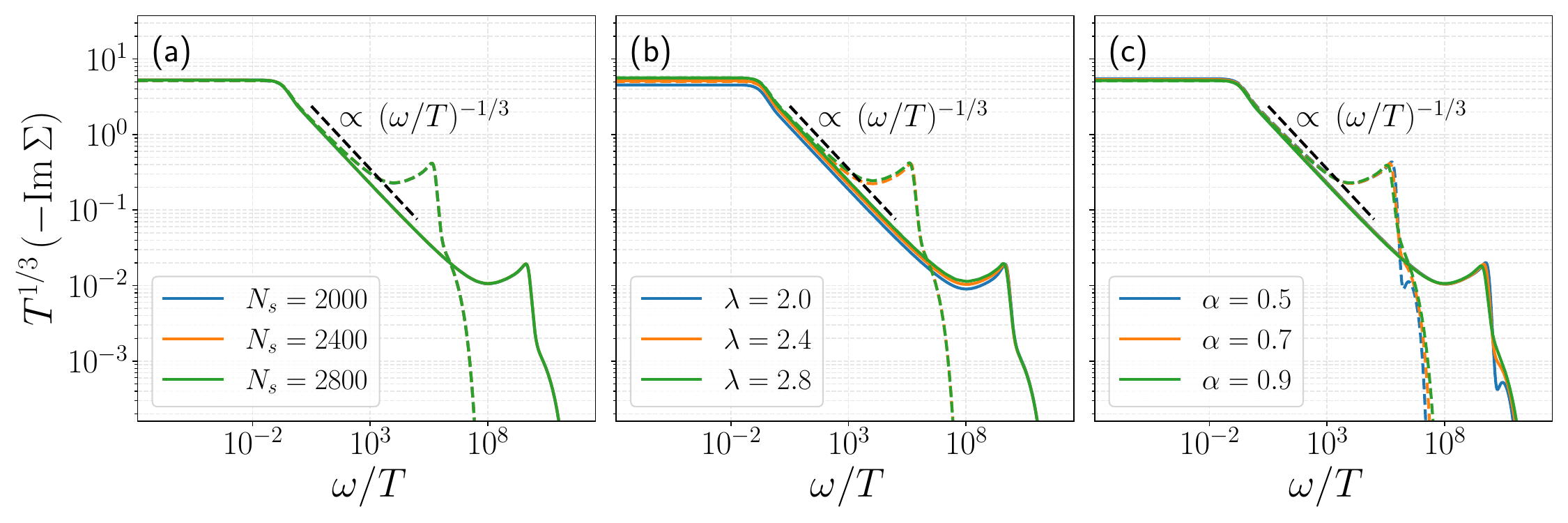}
    \caption{\textbf{NRG hyperparameter scan of the imaginary part of the electron self-energy.}  
    All panels show $\mathrm{Im}\,\Sigma(\omega,T)$ corresponding to the model parameters used in
    Fig.~\ref{fig:2}.  
    Solid lines denote $T=10^{-10}$ and dashed lines $T=10^{-6}$; data are plotted as 
    functions of $\omega/T$ and rescaled by $T^{1/3}$.  
    Different colors correspond to variations of the standard NRG hyperparameters:
    \textbf{(a)} number of kept states, 
    \textbf{(b)} discretization parameter $\Lambda$, and 
    \textbf{(c)} broadening parameter $\alpha$.  
    }
    \label{fig:A1}
\end{figure}

In this appendix we present a consistency check of the DMFT calculations used throughout the main text.  
Specifically, we analyze the convergence of the electron self-energy with respect to
the main NRG hyperparameters: the number of kept states $N_s$, the discretization parameter
$\Lambda$, and the broadening parameter $\alpha$.

We consider the same model parameters as in Fig.~\ref{fig:2} of the main text.
Figure~\ref{fig:A1} displays the imaginary part of the self-energy for two representative 
temperatures, $T=10^{-10}$ and $T=10^{-6}$, plotted as functions of $\omega/T$ and 
rescaled by $T^{1/3}$.
Each panel varies one NRG hyperparameter while keeping the others fixed, allowing us to
assess the numerical stability of the obtained scaling curves.

Panel~(a) shows the dependence on the number of kept states $N_s$ per iteration.  
Panel~(b) illustrates the effect of varying the discretization parameter $\Lambda$, which
controls the logarithmic sampling of energy shells.  
Panel~(c) examines the dependence on the broadening parameter 
$\alpha$ used in the full-density-matrix NRG (FD-NRG) spectral reconstruction.

While no $z$-averaging was used in panels~(a) and (b), we employed $N_z = 4$ in 
panel~(c) to ensure smooth, non-oscillatory spectra.  
This becomes essential when testing small broadening parameters, where 
discretization artifacts are otherwise more pronounced.

Across all panels, the results demonstrate excellent numerical stability:  
the self-energy curves collapse onto each other for all reasonable hyperparameter values,
confirming that the scaling behavior discussed in the main text is a robust physical feature
and not an artifact of the NRG setup.

\end{appendix}



\bibliography{References.bib}


\end{document}